\documentclass[aps,amsmath,amssymb, reprint, prl, preprintnumbers,superscriptaddress]{revtex4-1}

\usepackage{graphicx}%

\usepackage{hyperref} 

\usepackage{array}
\usepackage{longtable}
\usepackage{verbatim}

\newcommand{\beqa}{\begin{eqnarray}}
\newcommand{\eeqa}{\end{eqnarray}}

\newcommand{\bea}{\begin{eqnarray}}
\newcommand{\eea}{\end{eqnarray}}

\newcommand{\CN}{{\mathcal N}}
\newcommand{\CO}{{\mathcal O}}

\def\CO{{\cal O}}

\newcommand{\be}{\begin{eqnarray}}
\newcommand{\ee}{\end{eqnarray}}

\newcommand{\bn}{\begin{enumerate}}
\newcommand{\en}{\end{enumerate}}

\def\Tr{{\rm Tr}}

\begin{document}

\title{Large $N$ Gauge Theories with a Dense Spectrum and the Weak Gravity Conjecture}

\preprint{APCTP Pre2020-001, QMUL-PH-19-38}

\author{Prarit Agarwal}
\affiliation{CRST and School of Physics and Astronomy, Queen Mary University of London, London E1 4NS, United Kingdom}
\email{agarwalprarit@gmail.com}

\author{Jaewon Song}
\affiliation{School of Physics, Korea Institute for Advanced Study, 85 Hoegiro, Dongdaemun-gu, Seoul 02455, Korea}
\affiliation{Asia Pacific Center for Theoretical Physics, Pohang, Gyeongbuk 37673, Korea}
\affiliation{Department of Physics, Korea Advanced Institute for Science and Technology, 291 Daehak-ro, Yuseong-gu, Daejeon 34141, Korea}
\email{jaewon.song@kaist.ac.kr}

\begin{abstract}
We find large $N$ gauge theories containing a large number of operators within a band of low conformal dimensions. One of such examples is the four-dimensional $\CN=1$ supersymmetric $SU(N)$ gauge theory with one adjoint and a pair of fundamental/anti-fundamental chiral multiplets. This theory flows to a superconformal theory in the infrared upon a superpotential coupling with gauge singlets. The gap in the low-lying spectrum scales as $1/N$ and the central charges scale as $O(N^1)$ contrary to the usual $O(N^2)$ scaling of ordinary gauge theory coming from the matrix degree of freedom. We find the AdS version of the Weak Gravity Conjecture (WGC) holds for this theory, although it cannot be holographically dual to supergravity. This supports the validity of WGC in a more general theory of quantum gravity. 
\end{abstract}

\setcounter{tocdepth}{2}
\maketitle
\paragraph{Introduction}
The AdS/CFT correspondence provides a definition of quantum gravity in $(d+1)$-dimensional anti-de Sitter space via conformal field theory in $d$-dimensions \cite{Maldacena:1997re, Witten:1998qj, Gubser:1998bc}. The most well-studied examples of AdS/CFT typically involve supersymmetric gauge theories realized on the stack of $N$ branes in string/M-theory. The dual gravity description in bulk becomes semi-classical supergravity in the limit where string coupling goes to zero and the AdS radius becomes large. This is tantamount to taking large $N$ and large 't Hooft coupling limit in the boundary field theory side. 

However, the strong version of AdS/CFT asserts that the correspondence between AdS gravity and boundary CFT holds beyond the semi-classical/particle limit. It means that \emph{any} conformal field theory in $d$-dimensions is equivalent to a quantum gravity in $d+1$-dimensional AdS. For a finite $N$ theory (the parameter $N$ can be replaced by central charges in general even-dimensional CFTs) at generic coupling, the bulk description can be exotic (such as light string states, non-local interactions) and very different from Einstein gravity. 

One necessary condition for a large $N$ CFT to have a weakly-coupled Einstein-like holographic description in AdS is the sparseness of the low-lying spectrum \cite{Heemskerk:2009pn, ElShowk:2011ag}. It means that as we take the large $N$ limit, the gap between the low-lying operators scales as $\CO(1)$. This condition is also necessary for the confinement/deconfinement transition to occur, which is dual to the Hawking-Page phase transition \cite{Hawking:1982dh, Witten:1998zw}. 
The number of heavy ($\Delta \gg \CO(N^2)$) states grows exponentially, which is accounted by the black hole microstates. 

Typically, most large $N$ gauge theories in the 't Hooft limit satisfy this condition. This is because low-lying gauge-invariant operators are formed out of $\CO(N)$ elementary fields, such as $\Tr \Phi^i$ with $2 \le i \le N$ in the case of $\CN=4$ super Yang-Mills theory. The 't Hooft limit ensures that the possible anomalous dimensions for the elementary fields are under control. 
Therefore one natural question to ask is whether it is possible to have a large $N$ gauge theory that does not satisfy the sparseness condition, which is necessary (and maybe sufficient) to have a holographic description. 

In this paper, we show that there indeed exists large $N$ gauge theories with dense spectrum at low-energy. More precisely, the gap in the scaling dimensions for the low-lying operators scales as $1/N$, and the dimensions $\Delta$ of the `single-trace operators' lie within a band of $\Delta \in (1, 3]$. The central charges $a$ and $c$ of these theories grows linearly in the rank of gauge group $N$, contrary to the intuitive growth of matrix degrees of freedom $N^2$. 

The gauge theories we study turn out to be rather simple, but strongly-coupled and do not have any weak-coupling limit. Our theories contain a $U(1)_A \times U(1)_B$ flavor symmetry, and we test the AdS version of the Weak Gravity Conjecture (WGC) \cite{ArkaniHamed:2006dz, Nakayama:2015hga} for the charged states. We find the WGC holds for these theories even though they are not dual to semi-classical Einstein-like gravity.

\vspace{5pt}
\paragraph{The model: $SU(N)$ SYM theory with 1 adjoint and fundamental }

Let us consider the $\mathcal{N}=1$ supersymmetric $SU(N)$ gauge theory with 1 adjoint chiral multiplet $\Phi$ and a pair of fundamental/anti-fundamental chiral multiplets $(Q, \widetilde{Q} )$. 
Let us turn off any superpotential term. There are two flavor $U(1)$ symmetries that we call $U(1)_B$ and $U(1)_A$. 
The charge assignments for the various symmetries can be summarized in a table as follows:
\begin{align}
\begin{array}{c|c|c|c|c}
	& SU(N) & U(1)_B & U(1)_A & R \\
	\hline
Q & N & 1 & N & 1-N R_\Phi \\ 
\tilde{Q} & \bar{N} & -1 & N & 1-N R_\Phi \\
\Phi & \textrm{adj} & 0& -1 & R_\Phi 
\end{array}
\end{align}
The R-symmetry and $U(1)_A$ symmetry are subject to the anomaly constraint. 
To find the superconformal R-charge in the IR we have to invoke `$a$-maximization' \cite{Intriligator:2003jj}, which states that the correct IR R-charge maximizes the $a$-function. The central charges for 4d SCFT can be written in terms of trace anomalies \cite{Anselmi:1997am}:
\begin{align}
 a = \frac{3}{32} \left(3 \Tr R^3 - \Tr R\right), ~~ c = \frac{1}{32} \left(9\Tr R^3 - 5 \Tr R \right) 
\end{align}
Now the R-charge is fixed by evaluating $\frac{\partial a}{\partial R} = 0$, $\frac{\partial^2 a}{\partial R^2} < 0$. 

An additional caveat arises from the fact that all the operators must satisfy the unitarity constraint: Any gauge-invariant chiral operators should have a scaling dimension $\Delta$ greater than 1. During the course of $a$-maximization, it often happens that the resulting value of R-charges causes certain chiral operator dimensions to drop to 1 or lower. This indicates that the corresponding operator gets decoupled along the renormalization group flow. Its contribution to the $a$-function must then be removed, following which $a$-maximization has to be redone \cite{Kutasov:2003iy}. This cycle needs to be iterated over until no more operators decouple. One way to deal with the decoupled operator is to introduce a `flip field' $X_\CO$ for each would-be decoupled operator $\CO$ and add a superpotential term $W = X_\CO \CO$. The F-term for $X_\CO$ removes the free $\CO$ from the chiral ring \cite{Barnes:2004jj}. The role of the flip field has been investigated further in \cite{Benvenuti:2017lle, Maruyoshi:2018nod, Agarwal:2018oxb}, for example. 

\vspace{5pt}
\paragraph{A Band of Dense Spectrum}
Now, let us study the spectrum of this theory. The (single-trace) gauge-invariant operators of this theory are given as follows:
\begin{itemize}
	\item Coulomb branch operators: $\Phi^n, \ 2 \leq n \leq N$ 
	\item dressed mesons: $Q\Phi^n \widetilde{Q}, \ 0 \leq n \leq N-1 $
	\item `baryon': $ Q (\Phi Q) (\Phi^2 Q) \ldots (\Phi^{N-1} Q)$
	\item `anti-baryon': $ \widetilde{Q} (\Phi \widetilde{Q}) (\Phi^2 \widetilde{Q}) \ldots (\Phi^{N-1} \widetilde{Q})$
\end{itemize}
We suppressed the gauge indices in the above expression.  
Let us remark that the chiral operators charged under $U(1)_B$ (that we call baryon) have a very large engineering dimension. We now have to repeatedly $a$-maximize and remove gauge-invariant operators whose scaling dimensions fall below the unitarity bound $\Delta > 1$. 

We find that some of the Coulomb branch operators $\Phi^n$ with $n=2, 3, \ldots N$ get decoupled and are replaced by corresponding flip fields, but not all of them are decoupled for $N \ge 12$. Most of the dressed mesons remain coupled, but some of the low-lying ones hit the unitarity bound and get decoupled. We find none of the `baryons' decouple along the renormalization group flow. 

\begin{figure}[t]
 	\begin{center}
 		\includegraphics[width=3.4in]{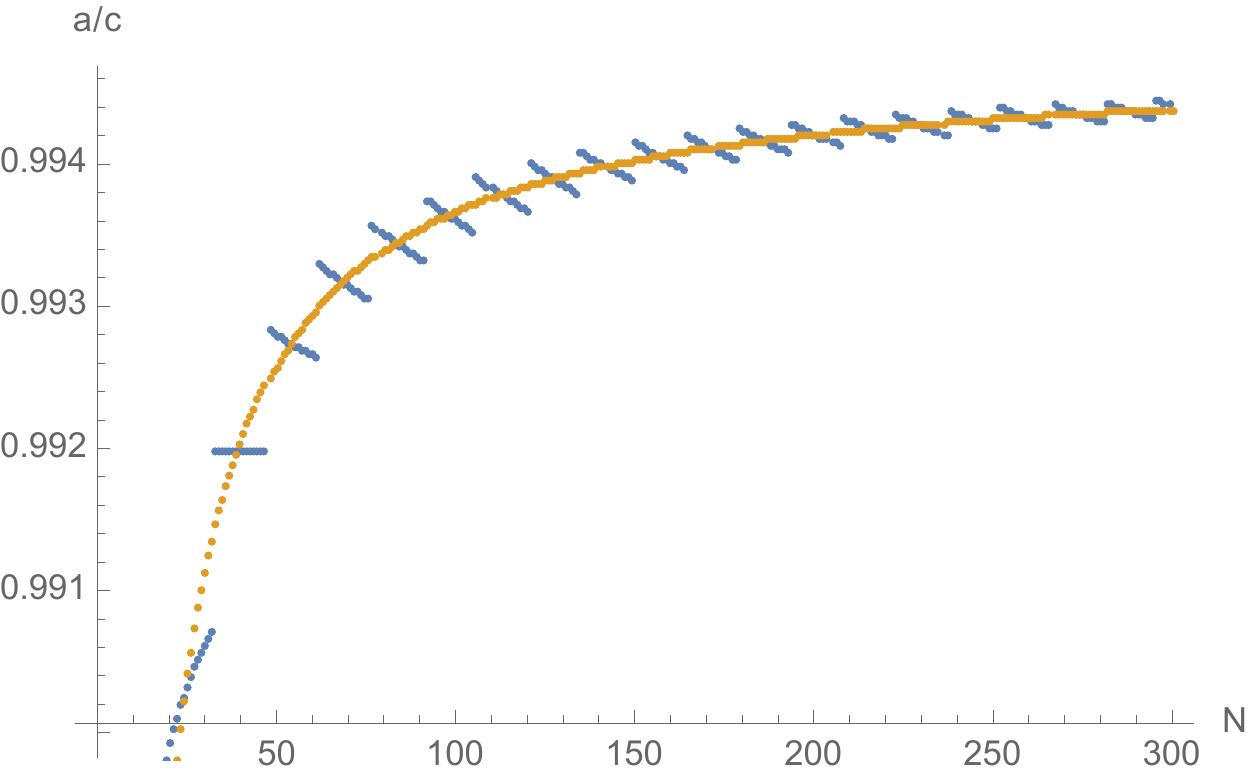}
 		\caption{Plot of $a/c$ vs $N$. The orange curve fits the plot with $a/c \simeq -0.1085/N + 0.9947$.}
 		\label{fig:AOverCVsNc}
 	\end{center}
\end{figure}

Due to the peculiarities arising from the pattern of decoupling of operators, it is somewhat technical to establish an analytical handle on the large-$N$ behavior of our theory. 
For now, we suffice ourselves with a numerical analysis of all gauge theories with $2 \leq N \leq 300$. Upon doing so, we obtain the IR central charges $a$, $c$ (upon removing the decoupled free chiral multiplets) behaves approximately as
\begin{align}
 a \simeq 0.5008 N - 0.6838 \ , \\
 c \simeq 0.5034 N - 0.6325 \ .  
\end{align}
We see that the central charges grow linearly in the rank of gauge group $N$, which is in stark contrast to the UV central charges given as $\CO(N^2)$. This is due to the very large quantum renormalization effect caused by the strong-coupling dynamics.  We plot the ratio of central charges $a/c$ as a function of $N$ in figure \ref{fig:AOverCVsNc}. Even though $a/c$ approaches close to 1, we find that it stays strictly lower than 1 for a sufficiently large $N$. 

\begin{figure}[h]
	\begin{center}
		\includegraphics[width=3.4in]{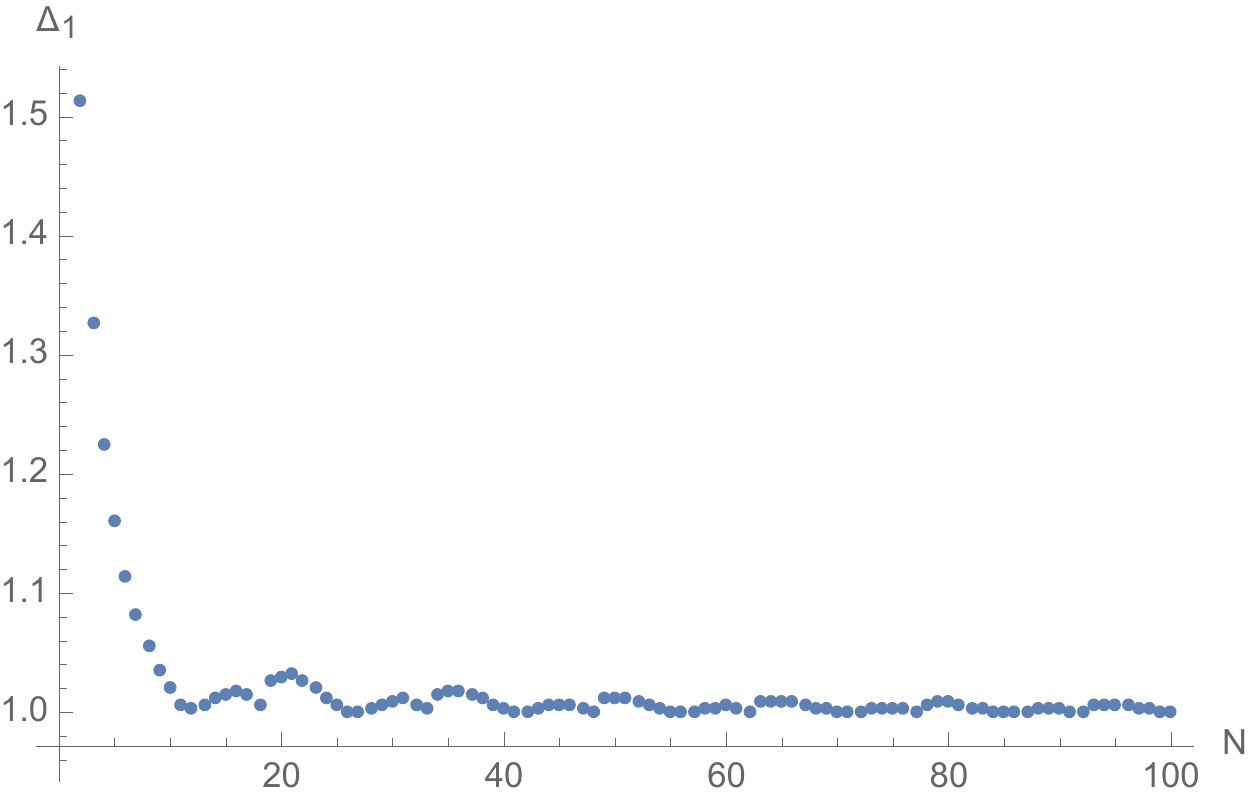}
		\caption{Plot of scaling dimension of the lightest operator $\Delta_1$ vs $N$}
		\label{fig:LightestVsNc}
	\end{center}
\end{figure}
The scaling dimension of the `lightest' operator $\Delta_1$ in the spectrum (the operator with the lowest scaling dimension) as a function of $N$ is depicted in figure \ref{fig:LightestVsNc}. The lightest operator turns out to be either given by the operator $Q \Phi^n \widetilde{Q}$ or $\Tr \Phi^n$ for some $n$ depending on $N$. For example, $Q\widetilde{Q}$ is the lightest operator for $2\le N \le 11$, $\Tr \Phi^N$ for $ 12 \le N \le 16$, and $Q\Phi \widetilde{Q}$ for $17 \le N \le 18$.  

\begin{figure}[h]
	\begin{center}
		\includegraphics[width=3.4in]{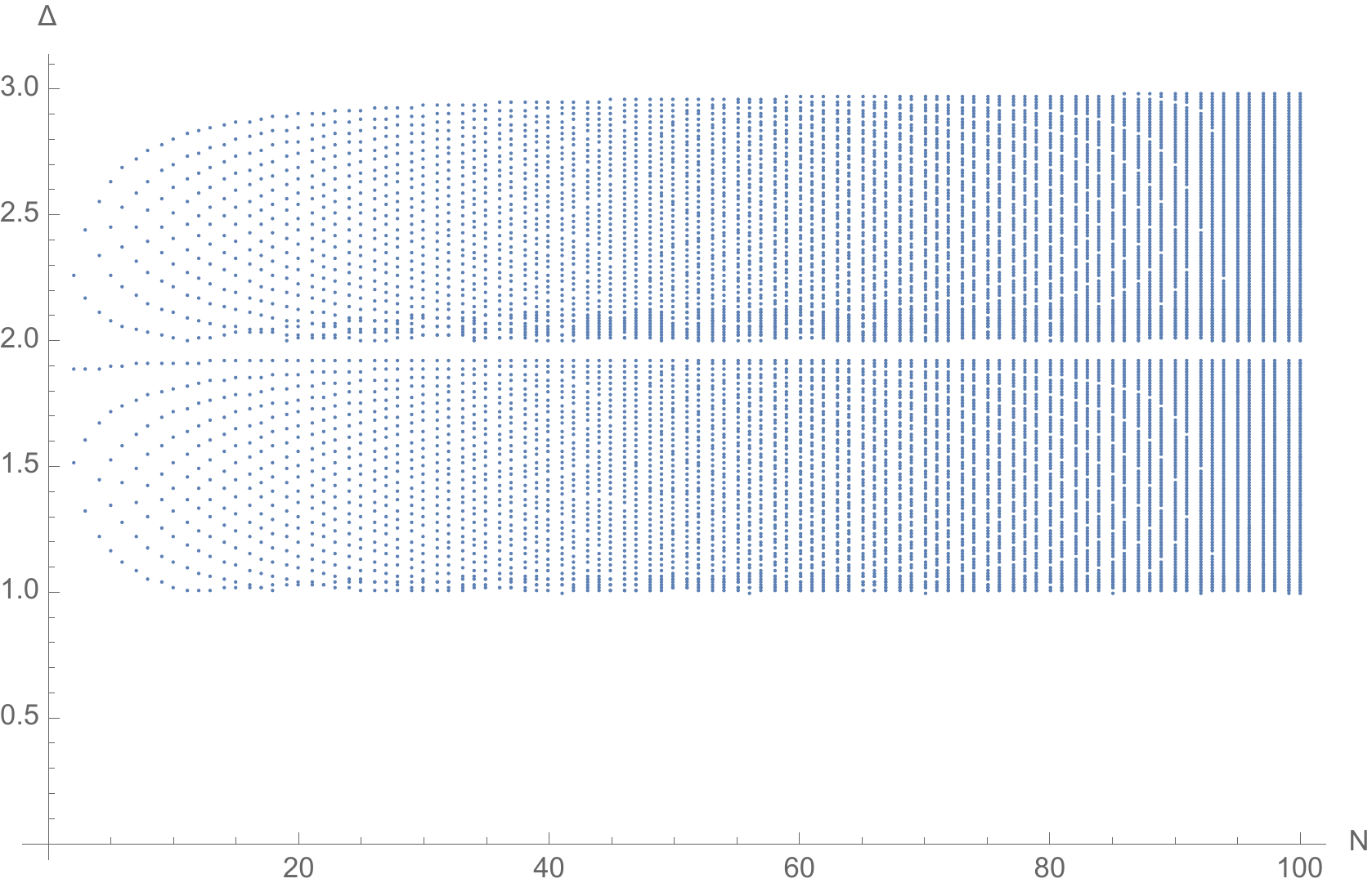}
		\caption{Dimensions of the `single-trace' operators form a band between 1 and 3.}
		\label{fig:spectrum}
	\end{center}
\end{figure}

We visualize the band of operator spectrum by plotting the dimensions of the chiral operators for each $N$ in figure \ref{fig:spectrum}. 
Here we do not include the baryon and anti-baryon operators, which are generally heavy ($\Delta_B \sim N$) and lie above the band. 
We find that the gap between the dimensions of the operators goes like $\CO(1/N)$. 
From this, we see that at large-$N$, the operator spectrum of the theory densely fills a narrow band of low conformal dimensions between 1 and 3.

\vspace{5pt}
\paragraph{Testing the Weak Gravity Conjecture} 

Given that the theory described above is not expected to have a weakly coupled gravity dual, it is interesting to check if its spectrum satisfies the bounds implied by the Weak Gravity Conjecture (WGC) \cite{ArkaniHamed:2006dz}. The implications of the WGC in AdS/CFT were proposed in \cite{Nakayama:2015hga}. There the authors argued that the WGC is essentially equivalent to requiring that extremal non-BPS Reissner-Nordstr\"om (RN) black holes in AdS must be unstable. Depending on the mechanism behind the decay of these black holes, they arrive at different bounds. The simplest of these comes from requiring that there should be a light charged particle to which even the smallest extremal AdS-RN black holes can decay. In the dual CFT, this implies that the spectrum should contain an operator with dimension $\Delta$ and charge $q$ such that
\begin{equation}
\label{eq:wgcBnd1}
\frac{\Delta^2}{q^2} \leq \frac{9 C_T}{40 C_V} \ ,
\end{equation}  
where $C_T$ and $C_V$ are the coefficients appearing in the two-point functions of the energy-momentum tensor and the corresponding flavor current, respectively. For a 4d $\CN=1$ SCFT, we can compute the coefficients by computing the trace anomalies as $C_T \sim \Tr R^3$ and $C_V \sim \Tr R F^2$ where $F$ is the flavor symmetry generator. 

There are two $U(1)$ flavor symmetries in our theory, that we label as $U(1)_A$ and $U(1)_B$. We test the WGC for these two flavor symmetries. 
We can estimate how the coefficients grow in $N$ using the trace anomaly formula and using the fact $R_\Phi \sim 1/N$ to get
\begin{align}
C_T &\sim \CO(N), \\
C_{V, A} &\sim  2N (R_Q-1)N^2 + N^2 (R_\Phi-1) \sim \CO(N^3) , \\
C_{V, B} &\sim 2N (R_Q-1) \sim \CO(N) . 
\end{align}
Therefore we obtain $\frac{C_T}{C_{V, A}} \sim \CO(1/N^2)$ and $\frac{C_T}{C_{V, B}} \sim \CO(1)$. 

\begin{figure}[h]
	\begin{center}
		\includegraphics[width=3.4in]{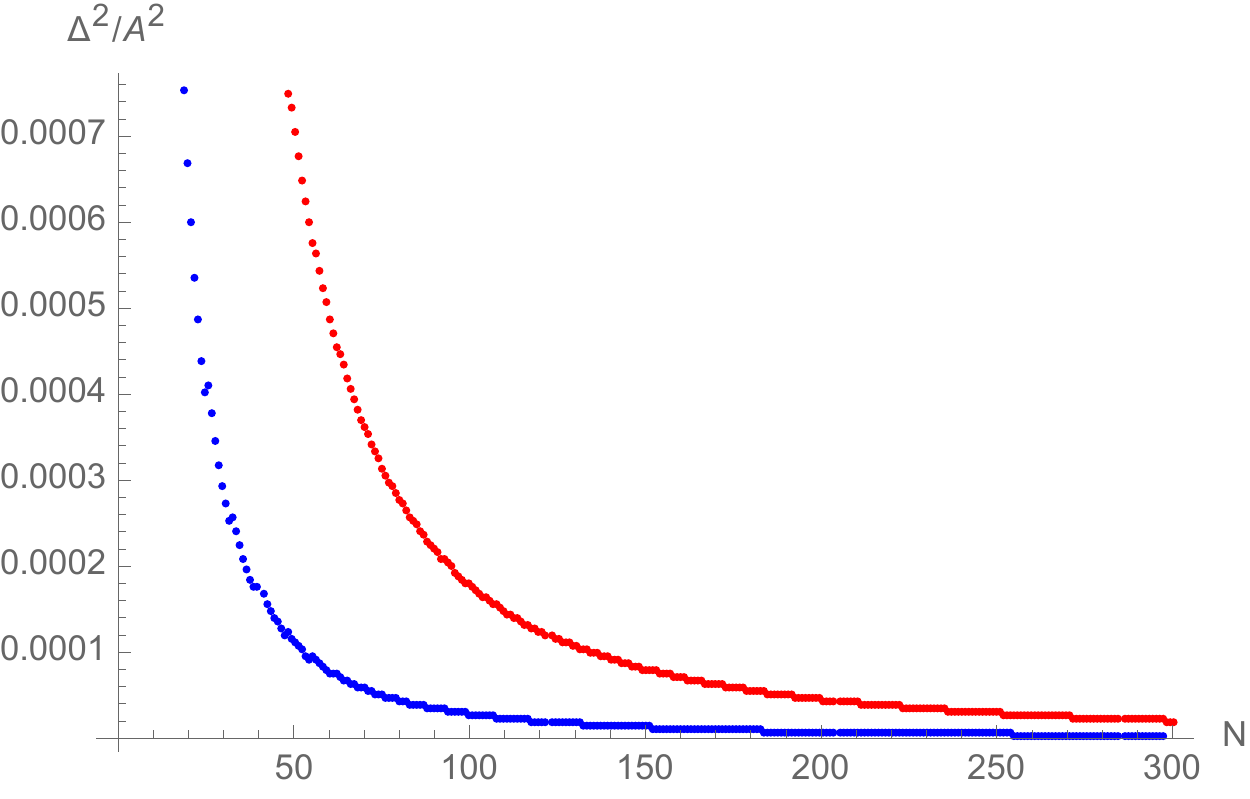}
		\caption{Test of the Weak Gravity Conjecture for $U(1)_A$. Red: $ 9 C_T/ 40 C_{V, A}$, Blue: $\Delta^2/q^2$ for the lightest meson. }
		\label{fig:wgc}
	\end{center}
\end{figure}

\begin{figure}[h]
	\begin{center}
		\includegraphics[width=3.4in]{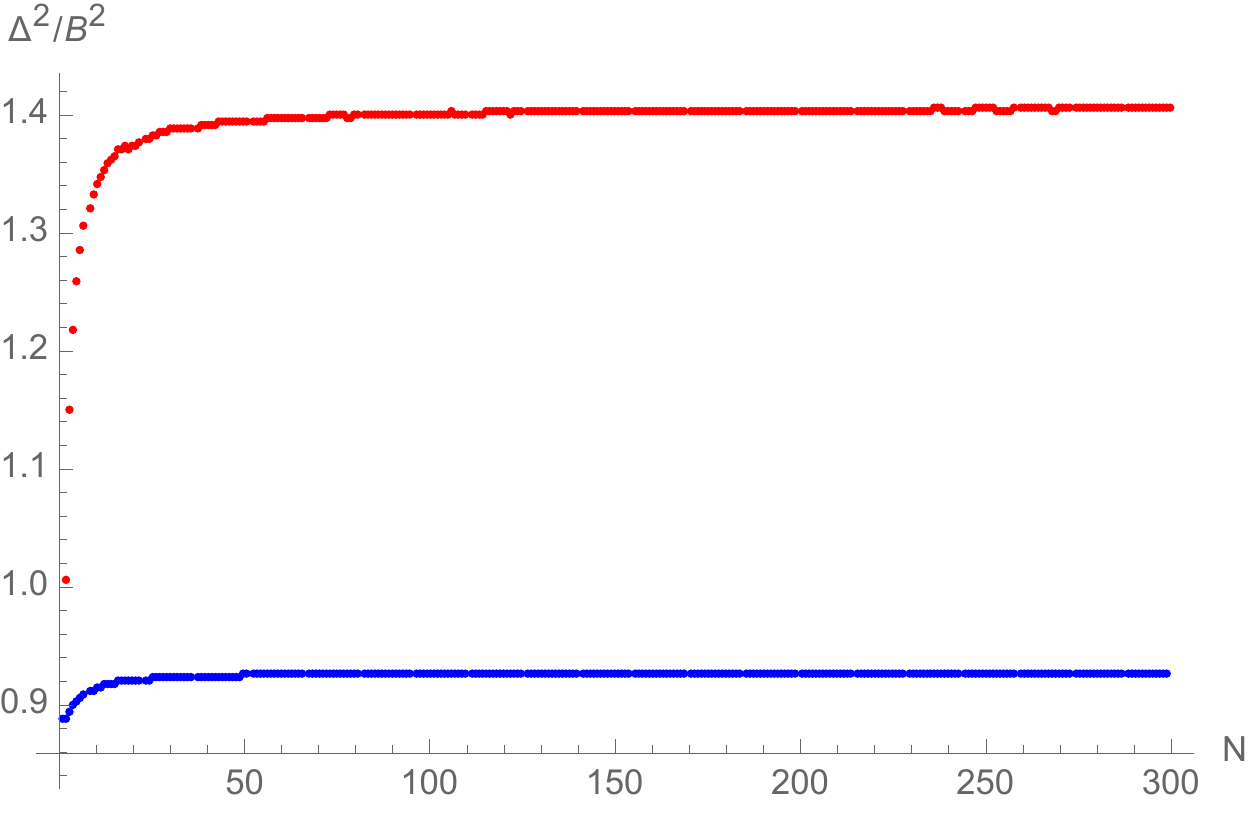}
		\caption{Test of the Weak Gravity Conjecture for $U(1)_B$. Red: $ 9 C_T/ 40 C_{V, B}$, Blue: $\Delta^2/q^2$ for the baryon operators charged under $U(1)_B$. }
		\label{fig:wgcb}
	\end{center}
\end{figure}

Let us check the WGC for the $U(1)_A$. 
Upon plotting $\Delta^2/q^2$ of the lightest mesonic operator $Q \Phi^n \tilde{Q}$ in the theory against the ratio $9 C_T/40 C_{V, A}$ in figure \ref{fig:wgc}, we see that indeed the WGC bound is satisfied. 
For the $U(1)_B$, we have `baryon' and `anti-baryon' with charge $B=\pm N$, and we plot the $\Delta^2/B^2$ against the ratio of $C_T$ and $C_{V, B}$ in figure \ref{fig:wgcb}. 
We also find that the WGC bound is satisfied by this operator. 

A second weaker bound was also obtained in \cite{Nakayama:2015hga} by considering the decay of AdS black holes through scalar hair formation. For small extremal black holes,  this gives
\begin{equation}
\label{eq:wgcBnd2}
\frac{(\Delta-2)^2}{q^2} \leq \frac{9 C_T}{40 C_V} \ ,
\end{equation}
which being a weaker bound is automatically satisfied in our theories. Slightly stronger bounds than \eqref{eq:wgcBnd1} are obtained by considering the formation of scalar-hair for intermediate $(r_+ \sim L_{AdS})$ and large $(r_+ \gg L_{AdS})$ sized black holes. The respective conditions are:
\begin{eqnarray}
\frac{4(\Delta-2)^2}{3q^2} \leq \frac{9 C_T}{40 C_V} \ , \\
\frac{3 (\Delta-1)(\Delta-3)}{ 2 q^2} \leq \frac{9 C_T}{40 C_V} \ .
\end{eqnarray}
We checked that these conditions are also satisfied by the lightest meson operator in our theory for the case of $U(1)_A$ and by the baryon operator in the case of $U(1)_B$. Therefore, it follows that our theory is indeed concordant with the Weak Gravity Conjecture despite it having a dense spectrum of light states. 

\vspace{5pt}
\paragraph{Another Model: $SU(N)$ theory flowing to the Argyres-Douglas theory}

Let us consider $SU(N)$ super Yang-Mills with a pair of fundamental/anti-fundamental chiral multiplets $(Q, \widetilde{Q} )$, a single adjoint chiral multiplet $\Phi$ and two set of $(N-1)$ gauge singlets $X_i, M_i$ coupled via superpotential couplings that flip all the Coulomb branch operators $\textrm{Tr} \Phi^{n+1}$, as well as the dressed mesons $\textrm{Tr}Q\Phi^{n-1} \widetilde{Q}$ for all $1 \leq n \leq N-1$. The superpotential is given by 
\begin{equation}
W = \sum_{i=1}^{N-1} X_i \textrm{Tr} \Phi^{i+1} + M_{i} \textrm{Tr} Q \Phi^{i-1} \widetilde{Q} .
\end{equation}
The spectrum of chiral operators (both in the UV and IR) in this case is simply given by $X_i, M_i \ \forall 1\leq i \leq N-1$ in addition to the operator $Q \Phi^{N-1} \widetilde{Q}$ which becomes the `moment-map' operator in the IR. 

It is known that this theory flows to the $(A_1, A_{2N-1})$ Argyres-Douglas (AD) theories \cite{Argyres:1995jj, Argyres:1995xn} at its IR fixed point \cite{Maruyoshi:2016tqk, Maruyoshi:2016aim}. We can therefore simply borrow the known results for AD theories to understand the growth of the central charges and operator scaling dimensions as a function of $N$. Thus the IR central charges are given by:
\begin{equation}
a = \frac{12 N^2-5N-5 }{24(N+1)}\ , \ c= \frac{3 N^2-N-1 }{6(N+1)} \ .
\end{equation}
As is the case of the previous example without flipping all the Coulomb branch operators and the dressed mesons, we see that the IR central charges grow linearly with $N$ such that $a, c \sim 0.5 N$ and $a/c \sim 1$ at large-$N$. The scaling dimensions of $M_i$ and $X_i$ are given by \cite{Fluder:2017oxm}
\begin{equation}
\Delta_{M_i} = \frac{2N-i+1}{N+1}, ~~ \Delta_{X_i} = \frac{3N-i+2}{N+1} \ , 
\end{equation}
with $i=1, \ldots, N-1$. A pair of $\CN=1$ chiral multiplets $(M_i, X_i)$ form an $\CN=2$ chiral multiplet, with the bottom component given by the scalar component of $M_i$. 

We see that the lightest chiral ring operator is given by $M_{N-1}$ while the heaviest chiral ring operator is given by $X_{1}$. As before, the scaling dimension of the lightest operator is only infinitesimally greater than 1 at large-$N$. Meanwhile, the gap in the scaling dimensions of the lightest and the second lightest operator decays as $\mathcal{O}(1/N)$, indicating the formation of a continuous band of operators with low conformal dimensions. We can also see this from the fact that the gap between the dimensions of the heaviest and lightest operator asymptotes to 2 while the number of chiral operators in the interacting sector grows linearly at large-$N$. 

The IR fixed point of this theory has enhanced $\CN=2$ supersymmetry and $U(1)_A \times U(1)_R$ symmetry gets enhanced to $SU(2)_R \times U(1)_r$. The $U(1)_B$ remains as the flavor symmetry of the theory with the moment-map operator given as $Z \equiv Q \Phi^{N-1} \widetilde{Q}$. If we denote the (anti)-baryon operator as $X \equiv Q^N \Phi^{N(N-1)/2}$ and $ Y \equiv \widetilde{Q}^N \Phi^{N(N-1)/2}$, there is a relation given by $X Y = Z^N$. They parametrize the Higgs branch of the theory, which is given by $\mathbb{C}^2/\mathbb{Z}_N$. 

The operator $X, Y$ has scaling dimension $N$ (which is consistent with the fact that $\Delta_Z=2$) and $U(1)_B$ charge $\pm N$. We find that the $\frac{9C_T}{40C_{V, B}} = \frac{3N^2 - N - 1}{2N^2} > 1$ in the large-N limit, whereas $\Delta^2/B^2 = 1$ for all $N$. Therefore the WGC is satisfied for the $U(1)_B$. 

\vspace{5pt}
\paragraph{Discussion}

In this paper, we showed that there exist large $N$ gauge theories with the dense low-lying spectrum, and the degrees of freedom measured by the central charges grow linearly in $N$. We focused on a set of theories given by 4d $\CN=1$ supersymmetric $SU(N)$ gauge theories. They provide interesting counter-examples of the commonly expected behavior of any large $N$ gauge theories, namely $\CO(N^2)$ degrees of freedom and sparse spectrum given by the gauge-invariant operators. Let us make a few comments regarding this model. 

As our models do not have a sparse low-lying spectrum, it is impossible to have a weakly-coupled Einstein dual in AdS$_5$. Nevertheless, our models satisfy the AdS version of the Weak Gravity Conjecture. This can be thought of as a piece of evidence that the WGC holds for more general quantum gravity, instead of being a special property of Einstein-like gravity. In the case of 2d CFT, modular invariance of the partition function implies the WGC \cite{Benjamin:2016fhe, Bae:2018qym, Aalsma:2019ryi}. It would be interesting to search for a condition in higher-dimension from which the WGC follows.

We expect there to be no confinement/deconfinement type phase transition at finite temperature or finite chemical potential. One way to probe the zero-temperature phase with varying chemical potential is by studying the superconformal index \cite{Kinney:2005ej, Romelsberger:2005eg}. Recently it was demonstrated that the index can indeed detect the phase structure, thereby accounting for the entropy of large AdS black holes \cite{Choi:2018hmj, Choi:2018vbz, Benini:2018ywd}. The Cardy-like formula for the 4d index \cite{Kim:2019yrz, Cabo-Bizet:2019osg} describing the high-energy behavior suggests that our model is unlikely to exhibit such a phase transition. However, it would be desirable to study detailed phase structure of our theory using the superconformal index, as was done in \cite{Choi:2018vbz, Cabo-Bizet:2019eaf, ArabiArdehali:2019orz}. 

Let us remark that there exists a `landscape' of large $N$ gauge theories that share the property (dense low-lying states) of our model. Many of the generalized Argyres-Douglas theories \cite{Cecotti:2010fi, Xie:2012hs} can be written in terms of gauge theories \cite{Agarwal:2016pjo, Agarwal:2017roi, Benvenuti:2017bpg}. Even with the same matter content, one can deform the theory via relevant operators and coupling with gauge singlets \cite{Maruyoshi:2018nod, MNS2}, which produce a large set of SCFTs sharing similar properties. 
It would be interesting to clarify the bulk interpretation of these theories.

\vspace{5pt}
\paragraph{Acknowledgments}
\begin{acknowledgments}
We would like to thank Dongmin Gang, Hee-Cheol Kim, Seok Kim, Ki-Hong Lee, Dario Rosa, and Shigeki Sugimoto for helpful discussion. We especially thank Ki-Hong Lee for pointing out an error in the previous version regarding the calculation of decoupling along the RG flow. 
The work of JS is supported by the Junior Research Group Program at the APCTP through the Science and Technology Promotion Fund, Lottery Fund of the Korean Government, Gyeongsangbuk-do, and Pohang City. 
The work of JS is also supported by the National Research Foundation of Korea (NRF) grant NRF-2020R1C1C1007591. 
The work of P.A. is supported by the Royal Society through a Research Fellows Enhancement Award, grant no. RGF\textbackslash EA\textbackslash 181049.
\end{acknowledgments}

\bibliography{refs}

\end{document}